\documentclass[a4paper,12pt]{article}
\usepackage{a4wide}
\usepackage[dvips]{graphicx,color}
\usepackage{amsmath,amsthm}
\usepackage{amssymb}

\newcommand{\nn}{\nonumber}

\newcommand{\tr}{\mathrm{tr}}
\newcommand{\Tr}{\mathrm{Tr}}

\renewcommand{\(}{\left(}
\renewcommand{\)}{\right)}
\renewcommand{\[}{\left[}
\renewcommand{\]}{\right]}

\allowdisplaybreaks[1]
\begin{document}

\begin{titlepage}
\begin{flushright}
LU TP 14-35\\
revised December 2014
\end{flushright}

\vfill
\begin{center}
{\large\bf LEADING LOGARITHMS FOR THE NUCLEON MASS}\\[1cm]
{\bf Johan Bijnens\footnote{E-mail: bijnens@thep.lu.se}
and Alexey A. Vladimirov\footnote{E-mail: vladimirov.aleksey@gmail.com}}\\[5mm]
Department of Astronomy and Theoretical Physics, Lund University,\\
S\"olvegatan 14A, SE 223 62 Lund, Sweden
\end{center}

\vfill

\begin{abstract}
Within the heavy baryon chiral perturbation theory approach,
we have studied the leading logarithm behaviour of the nucleon mass up to
four-loop order exactly
and we present some results up to six-loop order as well as
an all-order conjecture. The same methods allow to calculate the
main logarithm multiplying the terms with fractional powers of the quark mass.
We calculate thus the coefficients of $m^{2n+1}\log^{(n-1)}(\mu^2/m^2)$
and $m^{2n+2}\log^n(\mu^2/m^2)$, with $m$ the lowest-order pion mass.
A side result is the leading divergence for a general heavy baryon loop
integral.
\end{abstract}

\vfill

\end{titlepage}

\section{Introduction}

The calculation of high order terms in low-energy effective field theories
 (EFTs) is a difficult task. Nowadays, most interesting observables have
been calculated at the second order of the expansion, and the difficulty of 
these calculations shows little hope
for any further expansion. The main problem which restricts the potential of EFTs is their non-renormalizability. The non-renormalizability does
not bring any problem, in principle, for the calculation by means of counting schemes for EFTs, first introduced in \cite{Weinberg:1978kz}.
However, the rapidly increasing number of low-energy coupling constants (LECs), makes very high order applications practically of little use in
general.

Nevertheless, there are contributions of the higher order terms which are free from higher order LECs. In particular this is true for the
leading logarithmical (LLog) contributions. LLogs are not in general dominant for a generic observable. However, for some observables the LLog
contribution is dominant. Examples of such observables are the generalized parton distributions at small-$x$
\cite{Kivel:2007jj,Moiseeva:2013qoa} and certain $\pi\pi$ scattering lengths \cite{Colangelo:1995np}. In addition, the LLog terms are of great
theoretical interest because they allow us to judge the behaviour of a whole series of corrections in EFTs. We therefore
consider the calculation of LLog terms in EFTs as an interesting and useful task.

In renormalizable field theories the LLog terms can be calculated to all orders using the renormalization group (RG) and (simple) one-loop
calculations of the beta functions. In EFTs, as e.g. Chiral Perturbation Theory (ChPT), they can also be calculated using one-loop calculations
as was suggested already in \cite{Weinberg:1978kz}, and proven in \cite{Buchler:2003vw}. In contrast to renormalizable theories, in EFTs the
LLog terms cannot be obtained simultaneously for all orders, and every order of the perturbative expansion requires an additional
calculation. However, the evaluation of LLog terms is considerably simpler than a full calculation. As an example, the full two-loop leading
logarithms in bosonic ChPT were known long before the full results \cite{Bijnens:1998yu}.

Within bosonic EFTs the LLogs have been studied extensively. It has been shown that for EFTs with massless particles the LLog behaviour is
described by a closed set of equations with known kernels, which were elaborated in
\cite{Kivel:2008mf,Kivel:2009az,Koschinski:2010mr,Polyakov:2010pt}. Although, the analytical solution of these equations is not known, one can
generate numerically the first few hundreds coefficients rather fast, and use the approximate numerical solution in applications. An example is
the exploration of the ``chiral inflation'' of the pion radius within ChPT \cite{Perevalova:2011qi}. Taking into account the mass of the fields
allows for non-zero tadpole diagrams, which leads to a rapidly increasing number of equations with the chiral order since one has to consider
one-loop diagrams with an ever increasing number of external legs. Therefore, one needs to incorporate new processes at every new
order. As a result, the difficulty of the calculation grows extremely fast with the chiral order. By automatizing the procedure for a large
number of processes the LLogs are known up to seven loops for some quantities
\cite{Bijnens:2009zi,Bijnens:2010xg,Bijnens:2012hf,Bijnens:2013yca}.

The main goal of this paper is to generalize the methods used for bosonic EFTs
with masses to the nucleon case. As mentioned earlier, it
is not only interesting from the theoretical side, but also necessary for the
evaluation of nucleon parton distributions at $x\sim{m_\pi}/{M_N}$
\cite{Moiseeva:2013qoa,Moiseeva:2012zi}. In the paper we present the extension
of the RG method of \cite{Buchler:2003vw} to nucleon-pion ChPT. With its help,
we calculate the LLog coefficients for the chiral expansion of the nucleon mass
in the heavy-baryon formulation of ChPT. The main results are presented in
Sects.~\ref{sec:resultmass} and \ref{sec:conjectures}.
An earlier application of LLogs in the nucleon sector was the calculation of
the two-loop LLog contribution to the axial nucleon coupling constant $g_A$
\cite{Bernard:2006te}.

The paper is organized as follows: In Sect.~\ref{sec:RGO} we introduce the concept of renormalization group order (RGO). This is needed since in
the nucleon sector chiral counting and loop counting are not identical. Sect.~\ref{sec:meson} shows how the RGO concept
works in the meson sector and quotes some known results. Sect.~\ref{sec:HB_Lagrangian} introduces the heavy baryon ChPT Lagrangian in its two
most common variants and the different meson parametrizations we have used as a check on our result. Sect.~\ref{sec:general_comm} shows how the
RGO can be used to prove the calculation of the leading logarithms using only one-loop diagrams also in the nucleon sector. This is then used to
calculate the LLogs for the nucleon mass in Sect.~\ref{sec:nucl_mass}. Some technicalities are discussed in Sects.~\ref{sec:propagator} and
\ref{sec:pole}. We then calculate the LLogs for the nucleon mass as well as the odd-power next-to-leading logarithms (NLLogs) in
Sect.~\ref{sec:resultmass} up to four respectively five loops. The observed regularity in the leading logarithm allows to also calculate the
five loop result with a mild assumption. The LLogs, then essentially known to five loops, show a remarkable regularity when rewritten in the
physical pion mass. We conjecture that this regularity holds to all orders and in that case using the known results for the pion LLogs we have a
result for the nucleon mass LLogs up to 7 loops. This is described in Sect.~\ref{sec:conjectures}. A short numerical discussion of our results
is given in Sect.~\ref{sec:numerics}. We summarize our conclusions in Sect.~\ref{sec:conclusions}. The LLogs for a general heavy baryon one-loop
integral are discussed in App.~\ref{app:loopintegrals}.

\section{Renormalization group and order}
\label{sec:RGO}

\subsection{Renormalization group operator}

In this section we present a short, hopefully self-contained, introduction
to the renormalization group approach in EFTs. Our main goal is to
present the method of obtaining the dependence of observables on
the renormalization or subtraction scale $(\mu)$. The material is presented in
a form transparent for the application at higher orders. More extensive
discussions can be found in \cite{Buchler:2003vw,Bijnens:2009zi,AVthesis}.
In particular, what we call LLogs is the contribution with the highest power
of $\log\mu$ at a given order of the expansion.

To start with, we remind the reader that the Lagrangian of an EFT is the most
general local Lagrangian satisfying given symmetry properties with a given set of
degrees of freedom or fields. Such a Lagrangian contains an infinite number
of terms. In the absence of additional restrictions, every independent operator
is multiplied by an unknown coupling constant, usually called low-energy
constant (LEC).

It is convenient to multiply every operator by the counting parameter $\hbar$
to the power which reflects the minimal order of the perturbative
expansion the operator contributes to. In this way, the constant $\hbar$
resembles the coupling constant in a renormalizable field theory as a way to
keep track of (loop) orders in the expansion.
Therefore, an EFT Lagrangian takes the form
\begin{eqnarray}
\label{1:L=sum L^(n)}
\mathcal{L}^{\text{EFT}}_{\text{bare}}
=\sum_{n=0}^\infty \hbar^n\mathcal{L}^{(n)}_{\text{bare}}.
\end{eqnarray}
The Lagrangian $\mathcal{L}^{(n)}$ we call the Lagrangian of $n$'th
$\hbar$-order\footnote{The Lagrangian which contains the propagator of
fields, must be included in the zeroth $\hbar$-order.} and its LECs are
consequently called LECs of $n$'th $\hbar$-order. Let us, following
\cite{Bijnens:2009zi}, denote LECs of $n$'th $\hbar$-order as $c_i^{(n)}$,
where the index $i$ enumerates independent operators. In this way, the
$n$'th $\hbar$-order Lagrangian reads
\begin{eqnarray}
\mathcal{L}^{(n)}_{\text{bare}}=\sum_{i} c_{(\text{bare})i}^{(n)}\mathcal{O}^{(n)}_i.
\end{eqnarray}
For some low-energy EFTs, like mesonic ChPT, the $\hbar$-ordering of operators
is in one-to-one correspondence with the chiral ordering. However, the
definition (\ref{1:L=sum L^(n)}) is more general. It can be applied to any EFT,
and, even, to renormalizable theories, some examples can be
found in \cite{Buchler:2003vw,AVthesis}. We should mention that there is no
unique definition of the $\hbar$-ordering for a theory. The only
constraint is that the $\hbar$-order of an operator should increase with
increasing perturbative order. The choice made is for EFTs often referred to
as the choice of power counting.

The bare Lagrangian is now split into a part with
renormalized couplings $c_i^{(n)}$, which depend on $\mu$, and the counterterms.
The renormalization scale independence of the Lagrangian leads to the set of RG
equations for the LECs $c_i^{(n)}$. These equations are of the form
\begin{eqnarray}
\label{1:dc=beta}
\mu^2\frac{d}{d\mu^2}c_i^{(n)}(\mu^2)=\beta^{(n)}_{i}(\{c_j^{(m)}(\mu^2)\}),
\end{eqnarray}
where the beta-function is a polynomial in the LECs and we have indicated
explicitly the $\mu$-dependence. An important point, used later, is
that the right-hand side of (\ref{1:dc=beta}) contains only combinations of
LECs with total $\hbar$-order strictly less then $n$.

The general formal solution of the system of equations (\ref{1:dc=beta}) is
\begin{eqnarray}
\label{defR}
c_i^{(n)}(\mu^2)=\hat R\(\frac{\mu}{\mu_0}\)c_{i}^{(n)}(\mu^2_0)
=\exp\(\log\(\frac{\mu^2}{\mu^2_0}\)\hat H\)c_{i}^{(n)}(\mu^2_0).
\end{eqnarray}
This defines also $\hat R$. The operator $\hat H$ is defined as
\begin{eqnarray}
\label{defH}
\hat H= \int d\rho^2\sum_{n,i}\beta^{(n)}_i(\{c^{(m)}_j(\rho^2)\})
\frac{\delta}{\delta c_i^{(n)}(\rho^2)}.
\end{eqnarray}
The derivative in (\ref{defH}) is defined by
\begin{eqnarray}
\frac{\delta}{\delta c_i^{(n)}(\rho^2)} c_j^{(m)}(\mu^2)
= \delta_{ij}\delta^{mn}\delta\left(\rho^2-\mu^2\right)
\end{eqnarray}
such that
\begin{eqnarray}
\hat H c_i^{(n)}(\mu^2) = \beta_i^{(n)}(\{c_j^{(m)}(\mu^2)\})\,.
\end{eqnarray}
With the help of $\hat H$ or $\hat R$, one can obtain the coefficients of the
LLog for any observable, without actual calculation of loop
diagrams, if the beta-functions are already known. We will demonstrate this
explicitly in the next sections.

\subsection{Renormalization group order}
\label{sec:defRGO}

The crucial property of the operator $\hat H$ is that the repetitive action of $\hat H$ nullifies any given LEC (or products of LECs). This is
the direct consequence of two features of $\hbar$-counting. The first one is that the lowest order couplings, with $\hbar$-order equal to zero,
have zero beta-function, and therefore $\hat H c_i^{(0)}=0$. The second one is that the $\beta$-function of LEC $c_i^{(n)}$, as defined in
(\ref{1:dc=beta}), contains only products of couplings with total $\hbar$-order lower then $n$. Thus, every application of the operator
$\hat H$ onto a product of LECs lowers the total $\hbar$-order of that product, until it becomes zero.

For future convenience, we introduce the concept of renormalization group
order (RGO). A product $P_c$ of LECs has RGO $g$ if
\begin{eqnarray}
\hat H^g P_c \ne 0\quad\text{and}\quad\hat H^{g+1}P_c=0\,.
\end{eqnarray}
For a generic\footnote{We will use this term below to indicate that there are
exceptions where the beta-functions are zero ``accidentally.'' An
example of this is the constant $L_7^r$ in three-flavour bosonic ChPT. This
does not invalidate our later use of the RGO.} quantity with a tree
level contribution of $\hbar$-order $n$, the RGO is the same as the maximum
loop order that can appear when calculating that quantity to $\hbar^n$.

In the bosonic EFTs treated in the earlier works,
e.g.\cite{Kivel:2008mf,Bijnens:2009zi,Bijnens:2010xg,Bijnens:2012hf}, there is
a one-to-one correspondence between $\hbar$-order and RGO, namely $g=n$.
Therefore, the notion of RGO is unnecessary and was not used in these works.
However, such a relation does not hold in general, i.e. LECs of different
$\hbar$-order can have the same RGO. For example, in the nucleon ChPT,
the $\hbar$-counting as is related to RGO as $g=[n/2]$, where $[x]$ indicates
the integer part of $x$
(see detailed discussion in Sec. \ref{sec:general_comm}). In
such a case the use of the RGO concept is convenient.

It is natural to split the beta-function into terms with the same RGO. For
the beta-function of a coupling constant $c_i^{(n)}$ with RGO $g$ we can write
\begin{eqnarray}
\beta^{(n)}_{i}= \sum_{p=0}^{g-1}\beta^{(n,p)}_i(c).
\end{eqnarray}
Here $g=n$ or $[n/2]$ for the two cases mentioned above.
One can show that the part of the beta function with the highest RGO,
$\beta^{(n,g-1)}_i$, contains only contributions from one-loop. The next part,
$\beta^{(n,g-2)}_{i}$, contains contributions from one and two-loop diagrams
and so on. Thus, the expression for the operator $\hat H$ can be ordered by
RGO as
\begin{eqnarray}
\hat H=\sum_{p=1}^\infty \hat H_p\,.
\end{eqnarray}
$\hat H_p$ contains the beta-functions $\beta^{(n,g-p)}$ of the coupling
constants $c_i^{(n)}$ of RGO $g$. As a consequence,
acting with $\hat H_p$ on an expression reduces its RGO by $p$.

\section{LLog in mesonic ChPT}
\label{sec:meson}

In mesonic ChPT the choice of $\hbar$-counting versus chiral counting relates
both as $\hbar^n\sim \mathcal{O}(p^{2n+2})$. The
lowest order Lagrangian is of the second chiral order and
reads\footnote{We write here only the terms relevant for the mass and
neglect external fields.}
\begin{eqnarray}
\mathcal{L}^{(0)}_{\pi}=\frac{F^2}{4}\tr\[u_\mu u^\mu+\chi_+\],
\end{eqnarray}
where we use the standard notation
\begin{eqnarray}
\label{2:u_mu_def}
u_\mu=i\(u^\dagger \partial_\mu u-u\partial_\mu u^\dagger\),~~~~\chi_+=u^\dagger \chi u^\dagger+u\chi^\dagger u=m^2\(u^2+u^{\dagger2}\).
\end{eqnarray}
Here and though the text, $F$ is the bare pion decay constant and $m$ is the
bare pion mass, $m^2 = 2B\hat m$ in the notation of
\cite{Gasser:1983yg}. $u$ contains the meson fields, 
a few examples of possible parametrizations are given in
(\ref{u_param1}-\ref{u_param3}). The next order Lagrangian $\mathcal{L}^{(1)}$
is of fourth chiral order. The absence of odd chiral order
Lagrangians is guaranteed by Lorentz invariance.

In mesonic ChPT, the generic RGO of a LEC $c^{(n)}$ is equal to $n$. The one-to-one correspondence between generic RGO and the $\hbar$-order is
the result of the absence of odd-chiral-order Lagrangians. Any product of LECs is also in one-to-one correspondence with its generic RGO, which
is equal to the sum of the LECs' $\hbar$-orders. That, in turn, results in the simple ordering of beta-functions: the beta-function
$\beta^{(n,n-l)}$ contains only $l$-loop beta functions.

As an example of using the operator $\hat H$ of (\ref{defH}) to obtain the
LLog, we look at the physical pion mass. In order to obtain the
physical pion mass one should solve the equation
$m_\pi^2-m^2+\Sigma_\pi(m_\pi^2,m^2)=0$, where $\Sigma_\pi(p^2,m^2)$ is a series
of perturbative corrections to the pion propagator. The expression for
$\Sigma_\pi$ has the general form
\begin{eqnarray}
\label{2:Sigma_pi} \Sigma_\pi(p^2=m_\pi^2,m^2)
  = m^2\sum_{n=1}^\infty \(\frac{m^2}{(4\pi F)^2}\)^{n}
   \Sigma^{(n)}_\pi\(\frac{\mu^2}{m^2},c(\mu^2)\),
\end{eqnarray}
where $\Sigma^{(n)}_\pi$ is a dimensionless expression of maximum
$\hbar$-order $n$.
The first argument of $\Sigma_\pi^{(n)}$ appears only as
the argument of logarithms. We have suppressed the arguments $p^2/m^2$.
Note, that $p^2/m^2$ can also enter the arguments of logarithms,
moreover there can be logarithms of more complicated expressions of it.
Such logarithms are not RG logarithms, and can not in general be obtained by
any procedure based on RG.

The expression for $\Sigma_\pi$ is renormalization scale
independent\footnote{We use here a scheme where all one-particle irreducible
diagrams are made finite, otherwise one should apply the argument to a well
defined Green function of external currents.}.
Moreover, it is renormalization scale invariant at every chiral order
independently:
\begin{eqnarray}
\left[\mu^2\frac{\partial}{\partial\mu^2}
+\sum_{i,n}\beta^{(n)}_i\frac{\partial}{\partial c_i^{(n)}(\mu^2)}\right]
\Sigma^{(n)}_\pi\(\frac{\mu^2}{m^2},c(\mu^2)\)
=
\left[\mu^2\frac{\partial}{\partial\mu^2}
+\hat H\right]
\Sigma^{(n)}_\pi\(\frac{\mu^2}{m^2},c(\mu^2)\)
= 0\,.
\end{eqnarray}
Therefore, we again have as solution, similar to (\ref{defR}),
\begin{eqnarray}
\label{2:Sigma=R Sigma}
\Sigma^{(n)}_\pi\(\frac{\mu^2}{m^2},c(\mu^2)\)=\hat
R\(\frac{\mu^2}{\mu_0^2}\)\Sigma^{(n)}_\pi\(\frac{\mu_0^2}{m^2},c(\mu_0^2)\).
\end{eqnarray}
Choosing $\mu_0^2=m^2$, one neglects all the RG logarithms in
$\Sigma^{(n)}_\pi$ on the right-hand-side.
Thus, all RG logarithms are collected in
the action of the operator $\hat R$.

The expression for $\Sigma^{(n)}_\pi$ has the following form
\begin{eqnarray}
\Sigma^{(n)}_\pi=\sum_{i}\{c^{(n)}_{i}\}V^{(n)}_i+\text{terms with lower RGO},
\end{eqnarray}
where $\{c^{(n)}_{i}\}V^{(n)}_i$ form the tree contribution to $\Sigma^{(n)}$.
The symbol $\{c^{(n)}_{i}\}V^{(n)}_i$ denotes here the products of $c_j^{(m)}$
with the highest possible RGO for the product. The $V^{(n)}_i$ depend on
 $p^2/m^2$. The highest power of the RG logarithm
$\log\mu^2$ in the expression (\ref{2:Sigma=R Sigma}) accompanies the highest
power of $\hat H$, in $\hat R = \exp(\log(\mu^2/\mu_0)\hat H)$, which gives a
non-zero result acting on $\Sigma^{(n)}$. Since every action of $\hat H$
reduces the RGO of expression, the coefficient of the
LLog is
\begin{eqnarray}
\label{2:LL_mass_pi} \Sigma^{(n)}_\pi\(\frac{\mu^2}{m^2},c(\mu^2)\)= \frac{1}{n!}\log^{n}\(\frac{\mu^2}{m^2}\)\hat H_1^{n}
\sum_{i}c^{(n)}_iV^{(n)}_i+\mathcal{O}\(\text{NLLog}\),
\end{eqnarray}
where NLLog is the the acronym for the next-to-leading logarithms, and
hence $\mathcal{O}(\text{NLLog})$ denotes the part of expression without
LLog.

Therefore, constructing the higher chiral order Lagrangians,
and calculating the one-loop beta-functions of their LECs, one can obtain
the LL coefficients without actual calculation of multi-loop diagrams.
Moreover, the result is independent on the details of the higher order
Lagrangians, as long as they are sufficiently general for the process at
hand \cite{Bijnens:2009zi}.
Practically it is convenient to use a non-minimal Lagrangian generated
``on-the-fly'' by the counterterms to one-loop diagrams only.
This was the approach used
in \cite{Bijnens:2009zi,Bijnens:2010xg,Bijnens:2012hf,Bijnens:2013yca} for
the mesonic theory for several processes.

The solution of the pole equation gives us the expression for the physical
pion mass to LLog accuracy. It is known up to sixth power of
logarithms \cite{Bijnens:2012hf} and reads
\begin{eqnarray}
\label{2:mPhys}
m^2_\text{phys}=m^2\(
  1-\frac{1}{2}L+\frac{17}{8}L^2-\frac{103}{24}L^3+\frac{24367}{1152}L^4
  -\frac{8821}{144}L^5+\frac{1922964667}{6220800}L^6+\cdots\),
\end{eqnarray}
where
\begin{eqnarray}
\label{2:L_def}
L=\frac{m^2}{(4\pi F)^2}\log\(\frac{\mu^2}{m^2}\).
\end{eqnarray}

\section{Heavy baryon Lagrangian}
\label{sec:HB_Lagrangian}

The nucleon-meson ChPT in the naive form has the problem of the large nucleon
mass $M$. There are several ways of dealing with the presence
of this large scale, each with advantages and disadvantages. In this article,
we use the heavy baryon approach to meson-nucleon ChPT since in this approach
all scales that explicitly appear are soft and there are no divergences
nor $\mu$-dependence associated directly with the scale $M$.

For the LLog calculation, we have to determine the Lagrangians of zero RGO.
For the pion-nucleon system, these are Lagrangians of the first and
the second chiral orders. The first chiral order Lagrangian, neglecting terms
with external fields, reads
\begin{eqnarray}\label{3:L0}
\mathcal{L}^{(0)}_{N\pi}=\bar N\(i v^\mu D_\mu+g_A S^\mu u_\mu\)N,
\end{eqnarray}
where $S^\mu$ is a spin vector. We use the standard notation for the field
 combinations (see also the definitions in (\ref{2:u_mu_def})):
\begin{eqnarray}
D_\mu&=&\partial_\mu+\Gamma_\mu, \qquad u^2=U, \qquad
\Gamma_\mu= \frac{1}{2}\(u^\dagger \partial_\mu u+u \partial_\mu u^\dagger\).
\end{eqnarray}

The second order Lagrangian is sensitive to redefinitions of the nucleon
field. The most standard form of the second chiral order heavy
baryon Lagrangian reads \cite{Bernard:1992qa,Bernard:1995dp}
\begin{eqnarray}
\label{3:BKKM}
\mathcal{L}_{\pi N}^{(1)}&=&\bar N_v\Big[
 \frac{(v\cdot D)^2-D\cdot D-ig_A \{S\cdot D,v\cdot u\}}{2M}
 +c_1\tr\(\chi_+\)+\(c_2-\frac{g_A^2}{8M}\)(v\cdot u)^2
 \nn\\&&
 \hspace*{1cm}
  +c_3 u\cdot u+\(c_4+\frac{1}{4M}\)i\epsilon^{\mu\nu\rho\sigma}u_\mu u_\nu v_\rho
          S_\sigma \Big]N_v.
\end{eqnarray}
A different but equivalent version of the second order chiral Lagrangian is
given in \cite{Ecker:1995rk}, and it reads
\begin{eqnarray}\label{3:EM}
\mathcal{L}^{(1)}_{N\pi}&=&
\frac{1}{M}\bar N\Big[
 -\frac{1}{2}\(D_\mu D^\mu+i g_A\{S_\mu D^\mu,v_\nu u^\nu\}\)
 +A_1 \Tr\(u_\mu u^\mu\)+A_2 \Tr\((v_\mu u^\mu)^2\)
\nn\\&&
\hspace*{1cm}
  +A_3 \Tr\(\chi_+\)+A_5i\epsilon^{\mu\nu\rho\sigma}v_\mu S_\nu u_\rho u_\sigma
  \Big]N.
\end{eqnarray}
The relation between the LECs $A_i$ and $c_i$ is the following
\begin{eqnarray}
A_1=\frac{Mc_3}{2}+\frac{g_A^2}{16},~~A_2=\frac{Mc_2}{2}-\frac{g^2_A}{8},~~A_3=Mc_1,~~A_5=Mc_4+\frac{1-g_A^2}{4}.
\end{eqnarray}

Although the S-matrix elements are independent of the parametrization of
the nucleon field, the contributions of individual diagrams, and
expressions for the beta-functions are dependent on the field parametrization.
Therefore, the comparison of results for calculations performed
in different parameterizations is a very strong check of a calculation. The
calculations presented in the next sections have been done in both
parametrizations of the nucleon field. Additionally for further cross-checks,
we used different parameterizations of the pion field $u$. We have
used
\begin{eqnarray}\label{u_param1}
u=\exp\(i\frac{\pi^a\tau^a}{2F}\),
\end{eqnarray}
\begin{eqnarray}\label{u_param2}
u=\sqrt{1-\frac{\vec\pi^2}{4F^2}}+i\frac{\pi^a\tau^a}{2F},
\end{eqnarray}
and
\begin{eqnarray}\label{u_param3}
u=\sqrt{\frac{Y}{\sqrt{2}}}+i\frac{\pi^a\tau^a}{F}\sqrt{\frac{1}{2Y}}
\qquad\text{with}\qquad Y=1+\sqrt{1-\frac{\vec\pi^2}{F^2}}.
\end{eqnarray}

\section{LLog in nucleon-meson ChPT: general comments}
\label{sec:general_comm}

The consideration of the LLog behavior of nucleon-meson systems is similar to
meson systems, but with some additional features. The main additional feature
of meson-nucleon systems is the presence of operators with odd number of
derivatives. Therefore, the relation between $\hbar$-order and chiral
order is $\hbar^n\sim \mathcal{O}(p^{n+1})$ for the single-nucleon sector of
the ChPT Lagrangian, and $\hbar^n\sim \mathcal{O}(p^{n+2})$ for the meson
sector of EFT Lagrangian. At the same time, every loop increases the chiral
order by at least two. Therefore, the RGO is not in
one-to-one correspondence with $\hbar$-order. A LEC of $n$'th $\hbar$-order
$c^{(n)}$ has generically an RGO $\[\frac{n}{2}\]$. This has
important consequences in the RG and LLog structure of the theory.

\begin{figure}[tb!]
\centering
\includegraphics[width=0.5\textwidth]{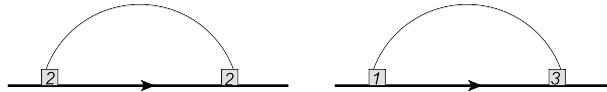}
\caption{An example of diagrams which contribute to the renormalization of the nucleon mass at the same chiral order (here, the fifth chiral
order), but which have different RGO. The left diagram has zero RGO, while the right diagram has RGO equal to one. Therefore, the left diagram
does not contribute to the LLogs. The number in the box indicates the chiral order of the vertex. Thick arrowed lines indicate nucleon
propagators, thin lines indicate pion propagators.}
\label{fig:1}
\end{figure}

The first consequence is the contribution of diagrams with different
RGO to the same chiral order. Indeed, a loop diagram with several
vertices of even chiral order (i.e. odd $\hbar$ order) has an RGO less
then a diagram with the same chiral order but with fewer
even-chiral-order vertices. An example of diagrams with the same chiral
order but different RGO is shown in Fig.~\ref{fig:1}.
Using the relation between chiral order and RGO, one can see that every
two even-chiral-order vertices reduce the RGO of a diagram by one, from
the possible maximum. For example: for any diagram with two even-chiral-order
vertices with certain RGOs, there exists diagrams of the same chiral order
and of the same topology, but with these two vertices replaced by
odd-chiral-order vertices, one with a chiral order one lower and one
chiral-order one higher. The lower chiral order has the same generic RGO as
the even vertex but the higher chiral order vertex has generic RGO one higher.
The latter diagram has thus a higher RGO. We conclude that at a given chiral
order the highest RGO contribution is given by the diagrams with
\textit{zero or one} vertex of even chiral order.

The second consequence is the ambiguity of the definition of a leading
logarithm. The natural definition of LLogs, the logarithm of a maximum
power of $\log\mu^2$ at a given chiral order, does not always coincides with
the RG definition for the same observable. In Sect.~\ref{sec:nucl_mass} we
will show that in the chiral expansion of the physical nucleon mass, the LLog
terms of odd chiral order are actually of NLLog origin when seen
from an RGO perspective.

\section{Nucleon mass at LLog accuracy}
\label{sec:nucl_mass}

\subsection{Propagator at LLog accuracy}
\label{sec:propagator}

The physical mass of the nucleon is given by  the position of the pole in the
Dyson propagator. In the heavy baryon approach the inverse Dyson
propagator reads (we remind the reader that superscripts $n$ refer to the
$\hbar$-order of quantities, and that the meson and meson-nucleon
sectors of the action have different chiral counting)
\begin{eqnarray}
\label{5:inverse_prop}
S^{-1}=(rv)+\sum_{n=1}^\infty \Sigma^{(n)}((r\cdot v),r^2),
\end{eqnarray}
where $r^\mu=p^\mu-Mv^\mu$, $p$ is the momentum of the nucleon, and $v_\mu$
is the reference four velocity of the heavy nucleon. There are some
intricacies of defining the propagator and renormalization of the nucleon
wave function in heavy baryon theory, see e.g.
\cite{Ecker:1995rk,Kambor:1998pi}. However for the determination of the
nucleon mass the straightforward usage of (\ref{5:inverse_prop}) is sufficient.

The expressions for $\Sigma^{(n)}$ are the result of the calculation of
one-particle irreducible diagrams with a nucleon line at the $(n+1)$'th
chiral order. The maximum power of the logarithm $\log\mu^2$ which can appear
in $\Sigma^{(n)}$ is $\[\frac{n}{2}\]$.

The derivation of the LLog coefficient is the same as the derivation of
expression (\ref{2:LL_mass_pi}). We collect all the RG-logarithms by the
action of the normalization point rescaling operator $\hat R$, again
suppressing the other arguments of $\Sigma^{(n)}$,
\begin{eqnarray}
\label{5:Sigma=R Sigma(nucl)}
\Sigma^{(n)}\(\frac{\mu^2}{m^2},c(\mu^2)\)=\hat
R\(\frac{\mu^2}{\mu_0^2}\)\Sigma^{(n)}\(\frac{\mu_0^2}{m^2},c(\mu_0^2)\).
\end{eqnarray}
The highest RGO in $\Sigma^{(n)}$ has the tree diagram with only a $c_i^{(n)}$
vertex, therefore, the LLog coefficient is given by
\begin{eqnarray}\label{5:LL_mass_nucl}
\text{LLog coef.}=
\(\[\frac{n}{2}\]!\)^{-1}\hat H_1^{\[n/2\]} \sum_{i}c^{(n)}_iV^{(n)}_i,
\end{eqnarray}
where $V^{(n)}_i$ are expression for the $c^{(n)}_i$-vertices.

It is also interesting to look at the NLLog contribution. The NLLog coefficient
comes from the $([n/2]-1)$ term of the exponent series in
(\ref{5:Sigma=R Sigma(nucl)}). There are several different parts
of $\Sigma^{(n)}$ which survive after the action by $\hat H^{[n/2]-1}$.
These are the terms with RGO $\[\frac{n}{2}\]$ and $\[\frac{n-2}{2}\]$. While
the first are given by tree diagrams, the second are given by
one-loop diagrams:
\begin{eqnarray}
\Sigma^{(n)}\(\frac{\mu^2}{m^2},c(\mu^2)\)=\sum_i c^{(n)}_i(\mu^2) V^{(n)}_i
+\sum_{\text{1-loop diag.}}W\(\frac{\mu^2}{m^2},c^{(k)}(\mu^2)\)+\cdots,
\end{eqnarray}
where the $V_i$ are the expressions for the tree level diagrams and $W$
indicates the  expressions for the one-loop diagrams at a fixed
renormalization scale. The dots represents the contributions with lower RGO.
We note that $W$ contains only one-loop diagrams with RGO equal to
$\[\frac{n-2}{2}\]$, but not all possible one-loop diagrams.
The NLLog coefficient is given by
\begin{eqnarray}
\label{5:NLL} \text{NLLog coef.}&=&
   \frac{1}{([n/2]-1)!}\hat H_1^{[n/2]-1}\sum_i c^{(n)}_i V^{(n)}_i
\nn\\&&
 +\frac{1}{([n/2]-1)!}\[\sum_{k=0}^{[n/2]-1}
     \hat H_1^{k}\hat H_2 \hat H_1^{[n/2]-1-k}\]\sum_i c^{(n)}_i V^{(n)}_i
\nn\\&&
  +\frac{1}{([n/2]-1)!}\hat H_1^{[n/2]-1} \sum_{\text{1-loop diag.}}
      W\(\frac{\mu_0^2}{m^2},c^{(k)}\),
\end{eqnarray}
where $\hat H_2$ contains two-loop beta-functions in addition to
one-loop beta-functions.

The expressions given by the different terms on the right-hand-side
in (\ref{5:NLL}) have significantly different properties. The first term
of (\ref{5:NLL}) gives the contribution to NLLogs with LECs from the
next-to-leading chiral Lagrangian. These terms are LLog terms of
$([n/2]-1)$-loop diagrams with insertion of higher order vertices.
The second line of (\ref{5:NLL}) gives the ``true'' NLLog contribution with
LECs of the lowest order Lagrangian only. These are the NLLog terms
of $[n/2]$-loop diagrams. The third line represents the non-analytical
contribution of $[n/2]$-loop diagrams to the NLLog coefficient. We should
mention that the part of NLLog coefficient given by the second line is
renormalization-scheme-dependent, while the parts given by the first and the
third lines are scheme-independent.

If the quantity has no tree-order contribution, the only non-zero part
of (\ref{5:NLL}) is the last line. In this case the NLLog can be
calculated from one-loop diagrams only. An example of such behavior
are the non-analytic in quark mass terms. These terms result only from
the loops and therefore, their contribution to NLLog can
be calculated with one-loop diagrams only.
The methods of \cite{Bijnens:2009zi} can also be used to prove this. The
absence of the tree level contribution allows the NLLog to be determined from
the set of equations relating the different loop-order contributions.

\subsection{Pole equation at LLog accuracy}
\label{sec:pole}

In this section we discuss the properties of the solution of the pole equation
at LLog accuracy. From the previous discussion it follows that it is
also valid for the NLLog multiplied by a nonanalytic power of the quark mass.

The position of the pole in the propagator (\ref{5:inverse_prop}) is a
Lorentz invariant quantity, when evaluated to all orders in the expansion.
Therefore, one can choose\footnote{If one chooses $v^\mu=(1,0,0,0)$
this corresponds to $\vec r=\vec p=0$.}
$r^\mu$ or $p^\mu$ such that $r^2=(r\cdot v)^2$.
Then, $\omega = (r\cdot v)$ gives the difference between
the physical mass and the bare mass, $(r\cdot v)=\delta M=M_{\text{phys}}-M$.
In this regime we can expand the expression for $\Sigma^{(n)}$ in
powers of $\delta M$
\begin{eqnarray}
\label{5:sigma_def}
\Sigma^{(n-1)}(\delta M)=\sum_{k=0}^n \sigma^{k,n-k}\delta M^k m^{n-k},
\end{eqnarray}
where the coefficients $\sigma^{(i,j)}$ contain logarithms. The coefficients
$\sigma^{k,n-k}$ have mass-dimension $(1-n)$.

The solution of the pole equation $S^{-1}=0$, (\ref{5:inverse_prop}),
can be found perturbatively in $m$
\begin{eqnarray}
\label{5:delta_expansion}
\delta M=\sum_{n=2}^\infty a_n m^n\,
\end{eqnarray}
where again the $a_i$ contain logarithms. Inserting the expansions (\ref{5:delta_expansion}) and (\ref{5:sigma_def}) into the Dyson propagator
(\ref{5:inverse_prop}), and considering the pole equation for every power of $m$ independently, we obtain a system of equations for the
coefficients $a_n$,
\begin{eqnarray}
\label{5:eqns}
a_{n}+\sum_{\{i\},j\leqslant n-2}a_{i_1}a_{i_2}...a_{i_j}\sigma^{j,n-\sum i}=0,
\end{eqnarray}
where summation runs over all possible sets of indices including empty set and permutations.

Let us consider the system of equations (\ref{5:eqns}) in the LLog regime.
We recall that the power of the LLog is $\[\frac{n}{2}\]$ for $\Sigma^{(n)}$.
However, the coefficients $a_n$ have different logarithm counting.
The reason is the presence of terms non-analytical in quark masses. The
Lagrangian of ChPT is necessarily analytical in the quark masses, i.e. it
contains only even powers of $m$. The terms non-analytic in quark masses
appear only through loop-integrals, and, therefore, they can not appear in
the expression (\ref{5:LL_mass_nucl}) or in the first two lines of the
expression (\ref{5:NLL}). In this way, the number of logarithms in front of
the pion mass in the odd power is suppressed by one (at least). Summarizing,
we obtain the following LLog counting for the coefficients $a$ and $\sigma$
\begin{eqnarray}\label{5:LL_counting}
 a_n \sim \log^{\[(n-2)/2\]}(\mu),~~~~\sigma^{s,t}\sim
 \left\{\begin{array}{cc} \log^{[(s+t-1)/2]}(\mu) & t\in\text{even}\\
                 \log^{[(s+t-3)/2]}(\mu) & t\in\text{odd}
        \end{array}
 \right. .
\end{eqnarray}

Using the counting (\ref{5:LL_counting}), we neglect the NLLog terms in the
equations (\ref{5:eqns}) and obtain the system of equations in the
LLog regime:
\begin{eqnarray}\label{5:a_even}
&&a_n+\sigma^{0,n}+\sum_{k=2,4,..}^{n-2}a_k \sigma^{1,n-k}=0,
\qquad n\in \text{even},
\\
\label{5:a_odd}
&& a_n+\sigma^{0,n}+a_{n-1}\sigma^{1,1}+\sum_{k=3,5,..}^{n-2}a_k \sigma^{1,n-k}=0,
\quad n\in\text{odd}.
\end{eqnarray}
This is a system of linear equations. The important result is that the
even-$n$ coefficients allow LLog evaluation only, because they
involve only the LLog coefficient of analytical in quark mass terms. At the
same time, the odd coefficients involve the terms non-analytical in
quark masses. These coefficients are really NLLog.. However, they can
be obtained from a one-loop calculation as well, because they follow from
the third line of (\ref{5:NLL}).

One can see that the system (\ref{5:a_even}-\ref{5:a_odd}) involves only the
coefficients $\sigma^{0,n}$ and $\sigma^{1,n}$, which are the
coefficients of the zeroth and the first powers of $(r\cdot v)$ in the
propagator diagrams. It is a reflection of the fact that according to
(\ref{5:LL_counting}) the quantity $(r\cdot v)^2=\delta M^2$ is of NLLog
order. Therefore, the powers of $\omega$ can be eliminated from the
equation $S^{-1}=0$. The solution can be presented in the simple form
\begin{eqnarray}\label{5:mass_sol}
\delta M= \frac{-\Sigma(0)}{1+\Sigma'(0)}+\mathcal{O}(\text{NLLog}),
\end{eqnarray}
where $\Sigma(r\cdot v)=\sum_{n}\Sigma^{(n)}(r\cdot v)$ and
$\Sigma'$ is its derivative with respect to $(r\cdot v)$.

\subsection{Expression for the physical mass}
\label{sec:resultmass}

We have performed the calculation of the nucleon mass up to the fourth power
of RG logarithms. We present the results in the form:
\begin{eqnarray}
\label{mainresult}
M_{\text{phys}}&=&M+k_2 \frac{m^2}{M}+k_3 \frac{\pi m^3}{(4\pi F)^2}
  +k_4 \frac{m^4}{(4\pi F)^2 M}\log\(\frac{\mu^2}{m^2}\)
  +k_5 \frac{\pi m^5}{(4\pi F)^4}\log\(\frac{\mu^2}{m^2}\)+\cdots
\nn\\
&=&
M+\frac{m^2}{M}\sum_{n=1}^\infty k_{2n} L^{n-1}
+\pi m\frac{m^2}{(4\pi F)^2}\sum_{n=1}^\infty k_{2n+1} L^{n-1},
\end{eqnarray}
where $L$ is defined in (\ref{2:L_def}). The coefficients up to $k_{11}$ are
presented in Tab.~\ref{table_mass}. This corresponds to the
four-loop calculation of LLog and five-loop calculation for the terms
non-analytical in quark masses.

The presented results have been obtained via the different parametrizations of the Lagrangians (see sec.\ref{sec:HB_Lagrangian}), which gives a
very strong check of calculation. Additionally, the coefficients up to $k_6$ agree with known results. The one-loop coefficients $k_{3,4}$ are
well known, see e.g. \cite{Bernard:1995dp}. The two-loop coefficient $k_{5}$ was first derived in \cite{McGovern:1998tm}. The two-loop
coefficients $k_6$ and $k_5$ are known from the full two-loop calculation for the nucleon mass performed in the EOMS scheme
\cite{Schindler:2007dr}.

\begin{table}[tbh!]
\begin{center}
\begin{tabular}{|c||l|}
\hline
\rule{0ex}{2.5ex} $k_2$ & $-4c_1 M$
\\[1mm]\hline
\rule{0ex}{2.5ex}$k_3$ & $-\frac{3}{2}g_A^2$
\\[1mm]\hline
\rule{0ex}{2.5ex}$k_4$ & $\frac{3}{4}\(g_A^2+(c_2+4c_3-4c_1)M\)-3c_1M $
\\[1mm]\hline
\rule{0ex}{2.7ex}$k_5$ & $\frac{3g_A^2}{8}\(3-16 g_A^2\)$
\\[1mm]\hline
\rule{0ex}{2.5ex}$k_6$ & $-\frac{3}{4}\(g_A^2+(c_2+4c_3-4c_1)M\)+\frac{3}{2}c_1M $
\\[1mm]\hline
\rule{0ex}{2.9ex}$k_7$ & $g_A^2\(-18 g_A^4+\frac{35 g_A^2}{4}-\frac{443}{64}\)$
\\[2mm]\hline
\rule{0ex}{2.5ex}$k_8$ & $\frac{27}{8}\(g_A^2+(c_2+4c_3-4 c_1)M\)-\frac{9}{2}c_1M $
\\[1mm]\hline
\rule{0ex}{2.9ex}$k_9$ & $\frac{g_A^2}{3}\(-116 g_A^6+\frac{2537 g_A^4}{20}-\frac{3569 g_A^2}{24}+\frac{55609}{1280}\)$
\\[2mm] \hline
\rule{0ex}{2.5ex}$k_{10}$ & $-\frac{257}{32}\(g_A^2+(c_2+4c_3-4c_1)M\)+\frac{257}{32}c_1M $
\\[1mm]\hline
\rule{0ex}{2.9ex}$k_{11}$ & $\frac{g_A^2}{2}\(-95 g_A^8+\frac{5187407 g_A^6}{20160}-\frac{449039
   g_A^4}{945}+\frac{16733923 g_A^2}{60480}-\frac{298785521}{1935360}\)$
\\[2mm]
\hline
\end{tabular}
\end{center}
\caption{The coefficients $k_i$ defined in (\ref{mainresult}) of the LLog expansion of the nucleon mass.} \label{table_mass}
\end{table}

The generation of the higher order Lagrangians and the evaluation of one-loop
beta-functions has been done automatically using the computer algebra system
FORM \cite{FORM}. The algorithm we used is similar to that used and described
in \cite{Bijnens:2009zi,Bijnens:2010xg}. The main integral needed for
evaluation of beta-functions is presented in the appendix. Although the
calculation involves only one-loop diagrams, it is very demanding in
machine time and memory. The most demanding factor is the length of the 
expression for the high order effective vertices and
the number of diagrams to compute.
These quantities grow rapidly with chiral order. For example, in order
to calculate the $k_{10}$ coefficient one needs to evaluate nearly $10^4$
one-loop diagrams.

The calculation of the even coefficients $k$ can be significantly simplified by using the conjectures discussed below in
Sect.~\ref{sec:conjectures}. So, by neglecting higher powers of $g_A$ during the evaluation of the diagrams, we could also evaluate the
five-loop coefficient $k_{12}$. Adding the further conjecture about the relation with the LLog in the pion mass, we can obtain the six and
seven-loop coefficients $k_{14}$ and $k_{16}$. However, these coefficients are the result of conjectures and, therefore, are presented in
Tab.~\ref{table_mass_conjecture}, separately from the results of the full calculation.

\subsection{Properties of the result and conjectures}
\label{sec:conjectures}

The straightforward calculation, limited by the available computer power,
gives us
the coefficients $k_1,\ldots,k_{11}$ presented in Tab.~\ref{table_mass}.
Considering the presented coefficients a number of regularities show up
immediately. Some of the regularities we can explain easily, while some of
them we cannot.

The first observation is that only even powers of $g_A$ show up.
This can be easily understood. The meson Lagrangian and the nucleon Lagrangian
are invariant under the transformation $u\leftrightarrow u^\dagger$
and $g_A\leftrightarrow -g_A$. As a consequence only terms even in $g_A$ can
appear in the nucleon mass. This explains the pattern occurring in the
odd coefficients $k_{2n+1}$.

The second observation is that the coefficients $k_{2n}$ contains a very
particular combination of LECs. The pattern appearing in $k_{2n}$ is not
well understood yet. Let us consider it in detail.

As shown in Sec.~\ref{sec:general_comm}, one can have at most one
insertion of the order $p^2$ Lagrangian. The expression for $k_{2n}$ are
thus at most linear in $c_1,c_2,c_3$ and the other terms
in $\mathcal{L}^{(1)}_{\pi N}$. The coupling constant $c_4$ or $A_5$ cannot
enter the nucleon mass at LLog since it produces
an $\epsilon_{\mu\nu\alpha\beta}$. However, we found no simple argument why
$g_A$ only appears up to order $g_A^2$.

For the powers of $g_A$,
there are two sources for factors of $g_A$ in the loop diagrams, namely
from the vertices $\mathcal{L}^{(0)}$ (\ref{3:L0}) and from the vertices
$\mathcal{L}^{(1)}$ (\ref{3:BKKM}-\ref{3:EM}). While the number of vertices
from $\mathcal{L}^{(1)}$ is restricted to one, the number of
vertices from $\mathcal{L}^{(0)}$ is naturally unrestricted. Moreover, the
expression for $\Sigma$ contains all allowed powers of $g_A$. These
powers cancels within the solution (\ref{5:mass_sol}). We have checked that if
one introduces new LECs for the terms proportional to
$g_A$ in $\mathcal{L}^{(1)}$ (say coefficients $B_{1,2}$ in front of the first
two terms in (\ref{3:EM})) the coefficients $k_{2n}$ would
contain higher powers of $g_A$. These induced higher powers are proportional
to $(B_1-B_2)$ and disappear when $B_1=B_2$. Since these operators
appear in $\mathcal{L}^{(1)}$ as the compensation of the non-relativistic
nucleon reference frame, we conclude that absence of higher powers
of $g_A$ in coefficients $k_{2n}$ is a consequence of Lorentz invariance.

Supposing that the cancelation of the higher powers of $g_A$ takes place at
all orders, one can neglect these powers during the computation of
diagrams. This procedure significantly reduces the demands for computer time
and allows us to calculate the coefficient $k_{12}$, which is
presented in Tab.~\ref{table_mass_conjecture}.

Considering the equation (\ref{5:a_even}) one can see that the coefficients
$k_{2n}$ consist from the terms proportional to exactly the first
power of $\sigma^{0,k}$ where $k$ is even. We remind that $\sigma^{0,\text{even}}$ are the result of diagrams with even chiral order and hence
proportional to a single vertex from $\mathcal{L}^{(1)}$ (which is also checked by explicit calculation with coefficients $B_{1,2}$). Therefore,
the term $g_A^2$ which appears in the coefficients $k_{2n}$ resulted solely from $\mathcal{L}^{(1)}$. In its own turn, it implies that there is
no contribution from the diagrams with vertices proportional to $g_A$ only from $\mathcal{L}^{(0)}$. All such vertices have an odd number of
pions. Absence of such vertices implies that \textit{diagrams with more than
 two odd-number-of-pion vertices do not contribute to the LLog
coefficient of nucleon mass}. Undoubtedly such a structure is a consequence
of the additional subtractions of infrared (heavy
mass) singularities into renormalization counterterms within heavy baryon
theory, but we have not been able to prove this.

Considering the first six coefficients $k_{2n}$ one can observe that they have
the pattern
\begin{eqnarray}
\label{5:pattern}
k_{2n}=b_n\(\frac{-3c_1M}{n-1}+\frac{3}{4}\(g_A^2+(c_2+4c_3-4c_1)M\)\),
\end{eqnarray}
where $b_n$ are some rational numbers. The coefficients $b_n$ can be obtained
from the calculation of the physical pion mass as we demonstrate
below.

The nucleon mass LLog coefficient in terms of the physical pion mass $m_{\text{phys}}$ has the form
\begin{eqnarray}
\label{mainresult2} M_{\text{phys}}&=&
M+\frac{m_\text{phys}^2}{M} \sum_{n=1}^\infty r_{2n} L_\pi^{n-1}
 +\pi m_\text{phys}\frac{m_\text{phys}^2}{(4\pi F)^2}
     \sum_{n=1}^\infty r_{2n+1} L_\pi^{n-1},
\end{eqnarray}
where
$$
L_\pi=\frac{m^2_\text{phys}}{(4\pi F)^2}\log\(\frac{\mu^2}{m_\text{phys}^2}\).
$$
The coefficients $r_n$ of this expansion are presented in
Tab.~\ref{table_mass_phys}.
\begin{table}[tb!]
\begin{center}
\begin{tabular}{|c||l|}
\hline
\rule{0ex}{2.5ex}$r_2$ & $-4c_1 M$
\\[1mm]\hline
\rule{0ex}{2.5ex}$r_3$ & $-\frac{3}{2}g_A^2$
\\[1mm]\hline
\rule{0ex}{2.5ex}$r_4$ & $\frac{3}{4}\(g_A^2+(c_2+4c_3-4c_1)M\)-5c_1M $
\\[1mm]\hline
\rule{0ex}{2.5ex}$r_5$ & $-6 g_A^4 $
\\[1mm]\hline
\rule{0ex}{2.5ex}$r_6$ & $5c_1M $
\\[1mm]\hline
\rule{0ex}{2.7ex}$r_7$ & $\frac{g_A^2}{4}\(-8+5g_A^2-72 g_A^4 \)$
\\[1mm]\hline
\rule{0ex}{2.5ex}$r_8$ & $\frac{25}{3}c_1M $
\\[1mm]\hline
\rule{0ex}{2.9ex}$r_9$ & $\frac{g_A^2}{3}\(-116 g_A^6+\frac{647 g_A^4}{20}-\frac{457 g_A^2}{12}+\frac{17}{40}\)$
\\[2mm]\hline
\rule{0ex}{2.5ex}$r_{10}$ & $\frac{725}{36}c_1M $
\\[1mm]\hline
\rule{0ex}{2.9ex}$r_{11}$ & $\frac{g_A^2}{2}\(-95 g_A^8+\frac{1679567 g_A^6}{20160}-\frac{451799 g_A^4}{3780}+\frac{320557
   g_A^2}{15120}-\frac{896467}{60480}\)$
\\[2mm]\hline
\end{tabular}
\end{center}
\caption{The coefficients $r$ of LLog expansion of the nucleon mass using the physical pion mass as defined in (\ref{mainresult2}).}
\label{table_mass_phys}
\end{table}

One can see that the non-analytical in quark mass terms $r_\text{odd}$ do not
simplify in this form of expansion, while the expressions for the
coefficient $r_\text{even}$ are significantly simplified. Moreover the
combination of the LECs proportional to $b_n$ in (\ref{5:pattern})
completely disappears from the higher order terms. We conclude that the
coefficients $b_n$ are the coefficients of the LLog expansion of $m^4$
in the terms of physical pion mass. Thus, assuming that the
 pattern~(\ref{5:pattern}) holds for all orders we conjecture the LLog part of
the expression for the nucleon bare mass via the physical
masses\footnote{This expression should be understood as not rewriting the term
 $k_2m^2/M$ in the physical pion mass and the integral
over $\mu^2$ should be done after applying (\ref{2:mPhys})
to $m^4_\text{phys}(\mu')$.} \textit{at all orders} to be
\begin{eqnarray}\label{6:M(m)}
M=
M_\text{phys} +\frac{3}{4}m_\text{phys}^4 \frac{\log\(\frac{\mu^2}{m_\text{phys}^2}\)}{(4\pi F)^2} \(\frac{g_A^2}{M_\text{phys}}-4c_1+c_2+4c_3\)
\qquad\qquad
\nn\\
 -\frac{3c_1}{{(4\pi F)^2}}\int\limits_{m^2_\text{phys}}^{\mu^2}m_\text{phys}^4(\mu')~\frac{d\mu'^2}{\mu'^2}.
\end{eqnarray}
The expression for the physical pion mass is known up to 6-loop order, Eq.~(\ref{2:mPhys}), therefore, we can guess two more LLog coefficients for
the physical nucleon mass. These are presented in the table \ref{table_mass_conjecture} and indicated by the double-star marks.

\begin{table}[tb!]
\begin{center}
\begin{tabular}{|c||l|}
\hline
\rule{0ex}{2.5ex}$k_{12}$(*) & $\frac{115}{3}\(g_A^2+(c_2+4c_3-4c_1)M\)-\frac{92}{3}c_1M $
\\[1mm]\hline
\rule{0ex}{2.5ex}$k_{14}$(**) & $-\frac{186515}{1536}\(g_A^2+(c_2+4c_3-4c_1)M\)+\frac{186515}{2304}c_1M $
\\[1mm]\hline
\rule{0ex}{2.5ex}$k_{16}$(**) & $\frac{153149887}{259200}\(g_A^2+(c_2+4c_3-4c_1)M\)-\frac{153149887}{453600}c_1M $
\\[1mm]\hline
\rule{0ex}{2.5ex}$r_{12}$(*) & $\frac{175}{4}c_1M $
\\[1mm]\hline
\rule{0ex}{2.5ex}$r_{14}$(**) & $\frac{4153903}{24300}c_1M $
\\[1mm]\hline
\end{tabular}
\end{center}
\caption{The coefficients $k_i$ and $r_i$ defined in (\ref{mainresult}) and
(\ref{mainresult2}), that are obtained by using the conjectures described in Sect.~\ref{sec:conjectures}. (*): The coefficients $k_{12}$ and $r_{12}$ have
been calculated within the simplified scheme by neglecting higher powers in $g_A$. (**):
The coefficients $k_{14,16}$ and $r_{14}$ are the result suggested by
the expression (\ref{6:M(m)}). $r_{16}$ would require the knowledge of the $L^7$ term in the expression for the pion mass.}
\label{table_mass_conjecture}
\end{table}

\subsection{Numerical results}
\label{sec:numerics}

\begin{figure}[tb!]
\centering
\includegraphics[width=0.5\textwidth]{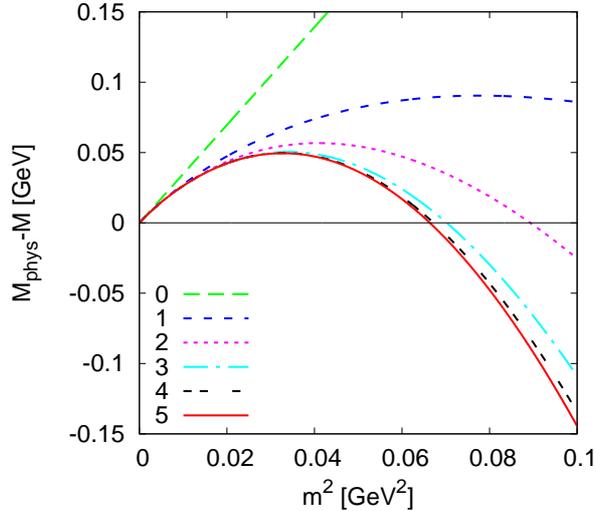}
\caption{The contribution of the terms in mass correction of (\ref{mainresult}) with the terms included up to a given loop-order.}
\label{figloop}
\end{figure}

As mentioned in the introduction, the LLog are not necessarily dominant.
They do however give an indication of the size of corrections to be expected.
We use here one set of inputs to show an example. The input we use uses
the $c_i$ as determined in \cite{Bernard:1995dp} and reasonable values for the
other quantities. The actual values we use are:
\begin{eqnarray}
M = 938~\text{MeV}, & c_1 = -0.87~\text{GeV}^{-1}, & c_2 = 3.34~\text{GeV}^{-1}, \quad \mu=0.77~\text{GeV},
\nn\\
F = 92.4~\text{MeV}, & c_3 = -5.25~\text{GeV}^{-1},& g_A= 1.25\,.
\end{eqnarray}
We plot in Fig.~\ref{figloop} the total correction $M_\text{phys}-M$
of (\ref{mainresult}) by loop order. We have included the results up to the
$k_{12}$ term since we do not have the odd powers higher than five loops.
As can be seen there is a reasonable convergence for the range given.

\begin{figure}[tb!]
\centering
\includegraphics[width=0.5\textwidth]{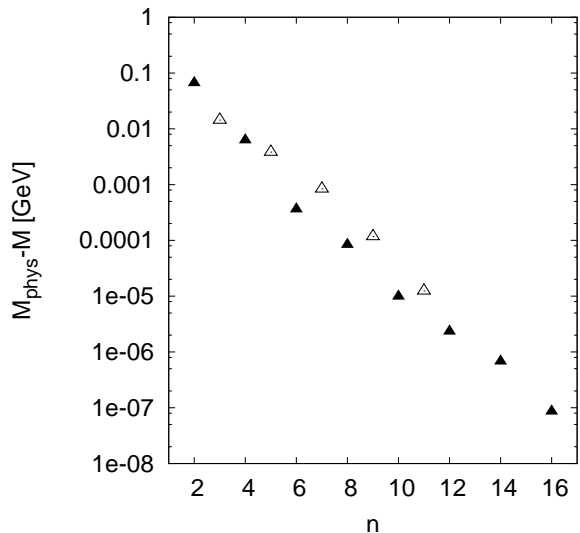}
\caption{The absolute value of the contribution of the individual terms ($\sim k_n$) in (\ref{mainresult}) at $m=138$~MeV. Open symbols are the odd orders. Filled symbols are the even orders.}
\label{figparts}
\end{figure}

To see the convergence better, we have plotted in Fig.\ref{figparts} the
absolute value of the individual terms containing $k_i$ of (\ref{mainresult})
for $m=138$~MeV.  Note the excellent convergence.

\section{Conclusions}
\label{sec:conclusions}

In this paper we have presented the application of the renormalization group
method for nucleon-pion chiral perturbation theory. The theoretical
basis of the method was developed in \cite{Buchler:2003vw}. The method has
been applied before only for bosonic theories, see
\cite{Kivel:2008mf,Polyakov:2010pt,Bijnens:2009zi,Bijnens:2010xg,Bijnens:2012hf,Bijnens:2013yca}. In particular, we have calculated the physical
mass of the nucleon within the heavy baryon formulation in the LLog
approximation (analytical and non-analytical in quark mass terms) up to
five-loop order. The results of the calculation are presented in
(\ref{mainresult}, \ref{mainresult2}) and tables
\ref{table_mass}, \ref{table_mass_phys}.

The theories with fermions (or more precisely, the theories involving
Lagrangians of odd chiral order) have a more involved
structure of RG equations. In contrast to the bosonic theories, where all
one-loop beta functions contribute to LLog coefficients, in theories
with fermions some one-loop beta functions do not contribute to the LLog
coefficients. For resolving the RG hierarchy, we have
introduced the concept of renormalization group order (RGO), see
Sec.~\ref{sec:defRGO}. Using the RGO allows us to extract the diagrams which
contribute to the LLog approximation (or any other order of RG logarithms).

The calculation of the necessary one-loop beta-functions has been performed
symbolically with the help of the FORM computer symbolic computation system.
The results of our calculation agree with known one- and two-loop results,
see e.g. \cite{Schindler:2007dr}. The calculations were performed in several
different parametrizations of the nucleon and pion fields with the same result,
providing a very strong check for the computational algorithm. The analytical
part of the computation, namely the expression for a basis one-loop-integral
in the heavy baryon theory, is presented in the appendix.

The obtained LLog coefficients show a number of regularities. Some of which
can be easily understood, while the rest is more involved. The most
intriguing regularity is the absence of higher powers of axial coupling
constant $g_A$ in the LLog coefficients $k_{2n}$ (see (\ref{mainresult})
and table \ref{table_mass}). Moreover, the pattern of the LLog coefficients
allows us to guess the \textit{all-order} expression for the LLog
contribution to the nucleon mass in terms of \textit{physical} pion mass 
(\ref{6:M(m)}). Although the latter is only a conjecture, we consider it
as an exact result, most likely a consequence of the subtraction of
heavy-mass-singularities within heavy baryon theory and Lorentz invariance.

We also showed some numerical results in Sect.~\ref{sec:numerics}.

\section*{Acknowledgements}

A.V. thanks J.~Relefors for stimulating discussions and technical help.
This work is supported in part by the European Community-Research
Infrastructure Integrating Activity Study of Strongly Interacting Matter"
(HadronPhysics3, Grant Agreement No. 28 3286) and the Swedish Research
Council grants 621-2011-5080 and 621-2013-4287.

\appendix

\section{Loop integrals}
\label{app:loopintegrals}

The most resource demanding part of LLog evaluation is the calculation of
the one-loop diagrams with generally a large number of external fields.
Also at high chiral orders the loop-integrals involves a large number of open
indices. In general in heavy baryon theory one faces
loop-integrals of the very general form
\begin{eqnarray}
\int\frac{d^dk}{(2\pi)^d}\frac{k^{\mu_1}..k^{\mu_n}}{(kv+\omega_1)...(kv+\omega_{N_N-1})((k+p_1)^2-m^2_\pi)...((k+p_{N_\pi+1})^2-m^2_\pi)},
\end{eqnarray}
where $N_N$ is the number of vertices with nucleon, and $N_\pi$ is the number
of pure pionic vertices involved in the diagram.

The first step of evaluation of such loop-integrals is the joining of
propagators of the same type into a single propagator with the help of
Feynman variables, $x_i$ for the nucleon propgators, $y_i$ for the meson
propagators. The subsequent shift of integration momentum allows one
to remove the momenta $p_i$ from the denominators (leaving them in
the ``mass''). The cost is a significant growth of the numerator, which is,
however, a purely algebraic problem.  The resulting sum integrals consist of
simpler base integrals, the expressions for which we present below. Finally,
the integrals over the Feynman parameters are done.

\subsection{The base mesonic integral}

The diagrams with a single or none nucleon vertex contain base integrals of
the form
\begin{eqnarray}
\label{app:int_base} I^{\mu\ldots\mu_n}_p=
 \int \frac{d^dk}{(2\pi)^d}\frac{k^{\mu_1}\ldots k^{\mu_n}}{(k^2-m^2)^{p}},
\end{eqnarray}
where $m^2$ is a difficult combination of $p_i$, Feynman parameters and
pion masses. The expression under the integral contains no intrinsic
vectors, the result is thus proportional to metric tensors only. Therefore,
the integral is zero for odd $n$. For even $n$ it is completely
symmetric in all the indices $\mu_i$ and reads
\begin{eqnarray}
\label{app:int_base_1}
I^{\mu\ldots\nu}_p= \frac{i(-1)^{\frac{n}{2}+p}}{(4\pi)^{\frac{d}{2}}}
 \frac{\Gamma\(p-\frac{n}{2}-\frac{d}{2}\)}{2^\frac{n}{2}\Gamma(p)}
g_s^{\mu_1\ldots\mu_n}(m^2)^{\frac{n}{2}-p+\frac{d}{2}},
\end{eqnarray}
where $g_s^{\mu\ldots\nu}$ is the totally symmetric combination of metric
tensors.
In our calculation we will need only the pole part of integral
$I$, it reads
\begin{eqnarray}
I^{\mu\ldots \nu}_p=\frac{i}{\epsilon(4\pi)^{\frac{d}{2}}} \frac{g_s^{\mu\ldots \nu}(m^2)^{2+\frac{n}{2}-p}}{2^\frac{n}{2}\Gamma(p)
\(2+\frac{n}{2}-p\)!} +\mathcal{O}(\epsilon^0),
\end{eqnarray}
where we have used that $d=4-2\epsilon$. This agrees with the expression
in \cite{Bijnens:2009zi}.

\subsection{The base integral with nucleon propagators}

The most general integral combines into the base integral of the form
\begin{eqnarray}
I^{\mu\ldots \nu}_{r,p}=   \int \frac{d^dk}{(2\pi)^d}
  \frac{k^{\mu}\ldots k^{\nu}}{(kv+\omega)^r(k^2-m^2)^{p}},
\end{eqnarray}
where $\omega=\sum_i \omega_i x_i$ (with $x_i$ Feynman parameters),
and $m^2$ is a combination of $p_i$, Feynman parameters $y_i$ and pion
mass. Using the pseudo-Feynman parameter $z$, we rewrite the integral as
\begin{eqnarray}
I^{\mu\ldots \nu}_{r,p}=2^r\frac{\Gamma(r+p)}{\Gamma(r)\Gamma(p)}
\int \frac{d^dk}{(2\pi)^d} \int_0^\infty dz
\frac{z^{r-1} k^{\mu}\ldots k^{\nu}}{(k^2-m^2+2z kv+2\omega z)^{r+p}},
\end{eqnarray}
where $z$ has mass dimension 1.

Performing the shift of the variable
\begin{eqnarray}\label{app:k->k+zv}
k^\mu ~\to~k^\mu-zv^\mu,
\end{eqnarray}
we represent the integral $I_{r,p}$ as a sum of (mesonic) base integrals
which can be evaulated with (\ref{app:int_base_1}). The resulting integral
over $z$ is of the form
\begin{eqnarray}
\tilde I&=& 2^r\frac{\Gamma(r+p)}{\Gamma(r)\Gamma(p)} \int_0^\infty dz
  \int\frac{d^dk}{(2\pi)^d} \frac{z^{r+l-1}k^{\mu_1}\ldots
   k^{\mu_n}}{\(k^2-m^2-z^2+2\omega z\)^{r+p}}
\nn\\\label{app:int_51}
 &=& \frac{i(-1)^{\frac{n}{2}-r-p}}{(4\pi)^\frac{d}{2}}
 \frac{g_s^{\mu..\nu}\Gamma\(r+p-\frac{n}{2}-\frac{d}{2}\)}
     {2^{\frac{n}{2}-r}\Gamma(r)\Gamma(p)}
\int_0^\infty dz~ z^{l+r-1}\(m^2+z^2-2\omega z\)^{\frac{n}{2}+\frac{d}{2}-r-p},
\end{eqnarray}
where $l\geq0$ follows from (\ref{app:k->k+zv}).

The parameters $p,r,n,l$ are all integers and satisfy $p,r\geq1$; $n,l\geq0$.
We introduce the special notation for the overall mass dimension of
the integral:
$$
A=n+l+4-r-2p\in\mathbb{Z}\,.
$$
In this notation the integral (\ref{app:int_51}) reads
\begin{eqnarray}\label{app:base_int_desected}
\tilde
I=\frac{i(-1)^{\frac{A-l-r}{2}}}{(4\pi)^\frac{d}{2}}\frac{g_s^{\mu..\nu}\Gamma\(\frac{r+l-A}{2}+\epsilon\)}{2^{\frac{n}{2}-r}\Gamma(r)\Gamma(p)}
\int_0^\infty dz~ z^{l+r-1}\(m^2+z^2-2\omega z\)^{\frac{A-l-r}{2}-\epsilon}.
\end{eqnarray}

We have two sources for the $\epsilon$-pole, namely, the $\Gamma$-function and
the integral over the pseudo-Feynman parameter $z$. Since
$l+r>0$, the only divergence in the $z$ integral takes place
at $z\to\infty$. We distinguish  three cases:
\begin{itemize}
\item[(i)] $A<0$~~the pseudo-Feynman integral is convergent.
Since $r>1$ and $l\geq0$, we have that $r+k-A\geq 1$. There is no pole
in $\epsilon$.
\item[(ii)] $A\geq0$~~the pseudo-Feynman integral is divergent, and
$r+l-A\geq 1$, such that the $\Gamma$-function has no pole.
\item[(iii)] $A\geq0$~~the pseudo-Feynman integral is divergent, but with
$r+l-A\leq 0$ such that the $\Gamma$-function also has a pole.
\end{itemize}
The important point is that all divergent cases are for non-negative $A$.
Expanding around $m=0$, and taking the integral for every term
we obtain
\begin{eqnarray}\label{app:base_int_111}
\tilde I&=&\frac{i(-1)^{\frac{3A-l-r}{2}-4\epsilon}}{(4\pi)^\frac{d}{2}}
  \frac{g_s^{\mu\ldots\nu}2^{A+r-\frac{n}{2}-2\epsilon}}{\Gamma(r)\Gamma(p)}
\\\nn&&
\times\sum_{j=0}^\infty \(-\frac{m^2}{4}\)^{j}
  \frac{\omega^{A-2 j-2\epsilon } \Gamma (2j-A+2\epsilon)
         \Gamma\(\frac{A+l+r}{2}-j-\epsilon \)}{j!},
\end{eqnarray}
The gamma-functions cannot have poles simultaneously, due to $l\geq1$.
We have two series of poles
\begin{eqnarray}
0\leq j\leq \frac{A}{2},~~~~\text{and}~~~~j\geq \frac{A+l+r}{2}.
\end{eqnarray}
The second series of poles appears at $z=0$. Therefore, these poles are
artifacts of the small-mass expansion. The first series
represents the UV-poles we want. Expanding the integral around these poles,
we obtain
\begin{eqnarray}
\label{app:base_int_epsilon_pole}
\tilde I=
\frac{i(-1)^{\frac{A-l-r}{2}}}{\epsilon(4\pi)^\frac{d}{2}}\frac{g_s^{\mu\ldots\nu}2^{A+r-\frac{n}{2}-1}}{\Gamma(r)\Gamma(p)}\sum_{j=0}^{A/2}
\(-\frac{m^2}{4}\)^{j}\frac{\omega^{A-2 j}\Gamma\(\frac{A+l+r}{2}-j\)}{j!(A-2j)!},
\end{eqnarray}
We note that this also implies that $A\geq 0$, and it is also the dimension
of the initial integral $I_{r,p}$. The argument of the last gamma-function is
always integer because $n$ is even.

\subsection{Non-analytical in quark mass part}

As shown in Sect.~\ref{sec:nucl_mass}, we could also consider the
terms non-analytical in the quark mass. They can be obtained from
(\ref{app:base_int_desected}). The integral over $z$ can be done exactly
with the help of Jacobi polynomials and then expanded in $\omega$. Or
it can be done in the following way. Expanding the integrand
of (\ref{app:base_int_desected}) in $\omega$ we obtain
\begin{eqnarray}
\nn
\tilde I= \sum_{j=0}^\infty
\frac{i(-1)^{\frac{A-l-r}{2}+j}}{(4\pi)^\frac{d}{2}}\frac{\omega^j
g_s^{\mu..\nu}\Gamma\(\frac{r+l-A}{2}+\epsilon\)}
                          {2^{\frac{n}{2}-r-j}\Gamma(r)\Gamma(p)}
\frac{\Gamma\(\frac{A-l-r}{2}-\epsilon+1\)}{j!\Gamma \(\frac{A-l-r}{2}-j-\epsilon+1\)}
\\
\times\int_0^\infty dz~ z^{l+r+j-1}\(m^2+z^2\)^{\frac{A-l-r}{2}-\epsilon-j}.
\end{eqnarray}
The integral over $z$ can be reduced to an Euler integral of the second kind.
After simplifications the result reads
\begin{eqnarray}
\tilde I=\frac{i(-1)^{\frac{A-l-r}{2}}}{(4\pi)^\frac{d}{2}}\frac{2^{r-\frac{n}{2}-1}g_s^{\mu..\nu}}{\Gamma(r)\Gamma(p)}\sum_{j=0}^\infty
(2\omega)^j m^{A-j-2\epsilon}\frac{\Gamma\(\frac{j+l+r}{2}\)\Gamma\(\frac{j-A}{2}+\epsilon\)}{j!},
\end{eqnarray}
which is equivalent to (\ref{app:base_int_111}).
Here we have three interesting cases:
\begin{enumerate}
\item $j\leq A$ (only for $A\geq0$), and $j-A$ even:  the expression contains
an $\epsilon$ pole, and is analytical in the quark mass. The
$\epsilon$-pole coefficient coincides with the one calculated before,
(\ref{app:base_int_epsilon_pole}).
\item $j>A$, and $j-A$ even: the expression is finite, and analytical in quark
mass. It is of no interest for this work.
\item $j-A$ odd: the expression is finite, but non-analytical in quark mass.
The leading term of the $\epsilon$-expansion reads
\begin{eqnarray}
\frac{i(-1)^{\frac{n}{2}-r-p}}{(4\pi)^\frac{d}{2}}\frac{2^{r-\frac{n}{2}-1}g_s^{\mu..\nu}}{\Gamma(r)\Gamma(p)}J_{n+l+4-r-2p}^{(l+r)}(\omega),
\end{eqnarray}
where
$$
J_A^{(s)}(\omega)=\sum_{j=0~(j+A~\text{odd})}^\infty (2\omega)^j m^{A-j}\frac{\Gamma\(\frac{j+s}{2}\)\Gamma\(\frac{j-A}{2}\)}{j!}\,.
$$
\end{enumerate}
The expressions for $A=2a$ (even) and $A=2a+1$ (odd) are:
$$
J^s_{2a}=2(-1)^a\sqrt{\pi}\frac{\Gamma\(\frac{1+s}{2}\)}{\(\frac{1}{2}\)_a}\frac{\omega}{m}(m^2)^a~_2F_1\(\frac{1+s}{2},\frac{1}{2}-a;\frac{3}{2};\frac{\omega^2}{m^2}\)\,,
$$
$$
J^s_{2a+1}=2(-1)^{a+1}\sqrt{\pi}\frac{\Gamma\(\frac{s}{2}\)}{\(\frac{3}{2}\)_a}m(m^2)^a~_2F_1\(\frac{s}{2},-\frac{1}{2}-a;\frac{1}{2};\frac{\omega^2}{m^2}\)\,.
$$

\end{document}